\documentclass[a4, 12pt]{article}

\usepackage{caption}
\usepackage{subcaption}
\usepackage{color}
\usepackage{eps fig}
\usepackage[svgnames]{xcolor} 

\usepackage{times} 

\usepackage{graphicx} 
\graphicspath{{figures/}} 
\usepackage{booktabs} 
\usepackage[font=small,labelfont=bf]{caption} 
\usepackage{amsfonts, amsmath, amsthm, amssymb} 
\usepackage{wrapfig} 
\usepackage[margin=1.5in]{geometry}
\usepackage{natbib}
\usepackage[bookmarks=false]{hyperref}

\begin{document}
\UseRawInputEncoding
\title{A Molecular Field Approach to Pressure Induced
Phase Transitions in Liquid Crystals: Smectic-Nematic transition}

\author{Sabana Shabnam$^{1}$, Sudeshna DasGupta$^{2}$, \\
Nababrata Ghoshal $^{3}$, Ananda DasGupta$^{4}$ and Soumen Kumar
Roy$^{5}$.}
\maketitle

$^{1}$School of Physical Sciences, National Institute of Science
Education and Research, HBNI, P. O. Jatni, 752050, India.\\
$^{1}$Email ID: shabnam.sabana@gmail.com, Phone No. 8981479826

$^{2}$Department of Physics, Lady Brabourne College, Kolkata 700017,
India. \\
$^{2}$Email ID: sudeshna.dasgupta10@gmail.com, Phone No. 983146102

$^{3}$Department of Physics, Mahishadal Raj College, Mahishadal,
Purba Medinipur, India.\\
$^{3}$Email ID: ghoshaln@yahoo.co.in, Phone No. 8335950698

$^{4}$Department of Physical Sciences, IISER Kolkata, Mohanpur 741246,
India.\\
$^{4}$Email ID: adg@iiserkol.ac.in, Phone No. 9831937272

$^{5}$Retd. Professor, Department of Physics, Jadavpur University,
Kolkata700032, India.\\
$^{5}$Email ID: roy.soumenkumar@gmail.com, Phone No. 9874741525.

\begin{abstract}
Since a rigorous microscopic treatment of a nematic fluid system based
on a pairwise interaction potential is immensely complex we had introduced
a simple mean field potential which was a modification of the Maier-Saupe
potential in a previous paper \cite{dasgupta2018pressure}. Building up on that here we have modified
that potential to take into account the various aspects of a smectic
A-nematic phase transition. In particular we have studied the dependence
of the phase transition on the coupling coefficient between the nematic
and smectic order parameters which in turn depends on the length of
alkyl chain, existence of tricritical point, variation of entropy
and specific heat as well as the dependence of the phase transition
on pressure.
\end{abstract}

\section{Introduction}

Maier and Saupe \cite{maier1960simple} presented a microscopic statistical
theory of nematic liquid crystals based on dipole-dipole dispersive
interactions. They obtained an orientation dependent potential between
a pair of elongated molecules which successfully described the first
order nematic-isotropic liquid phase transition in a molecular field
approximation. In this theory only the orientational order parameter
was considered. In 1970, Kobayashi \cite{k1970theory, kobayashi1971theory} introduced a
theory of melting in liquid crystals for which both the translational
and orientational order parameters were taken into account. His formulation
was analogous to that of Kirkwood and Monroe \cite{kirkwood1941statistical}. Later
in 1971, McMillan \cite{mcmillan1971simple,mcmillan1972x} presented a simple
molecular model with anisotropic forces for the smectic A phase by
extending the Maier-Saupe molecular potential. Introducing one physical
parameter $\alpha$ as a dimensionless interaction strength for the
smectic A phase, he predicted that the extent of the nematic range
would depend on the value of $\alpha$. The nematic range decreases
as $\alpha$ increases leading to a first order smectic A-isotropic
transition. One gets a triple point where the three phases, namely,
smectic A (A), nematic (N) and isotropic (I) coexist. McMillan
calculated the transition temperatures as a function of $\alpha$
and produced a generalized phase diagram to explain the role of alkyl
end chains in the formation of the smectic A phase. He was the first
to point out that the A-N transition could be second order if
$\frac{T_{AN}}{T_{NI}}<0.87$, where $T_{AN}$ is the A-N and
$T_{NI}$ the N-I transition temperatures. Above this value
the A-N transition is a first order transition. A change in
the order of phase transition is thus expected near $\frac{T_{AN}}{T_{NI}}=0.87$
giving rise to a tricritical point (TCP). According to this theory
the appearance of a second order phase transition is related to the
saturation of the nematic order at $T_{AN}$. McMillan's
prediction of second order behavior was subsequently verified experimentally
by Doane et al \cite{doane1972possible} in 1972. With the NMR study of the homologous
series of compunds like $4$-n-alkoxybenzylidene-$4^{\prime}$-phenylazoaniline,
they found that the A-N transition could be second order or
very nearly second order. The same qualitative features
were presented in the de Gennes \cite{de1973some} model. Using
a Landau expansion of smectic A free energy, he showed
that the order of the transition depends on $\alpha$ which is the
strength of the coupling between the nematic or orientational order
parameter and the smectic A or density order parameter. The alkyl chain length is related to the coupling
constant $\alpha$. As $\alpha$ increases which in turn increases
the coupling between the nematic and smectic A order parameters, the
extent of the nematic phase gradually becomes smaller. This favours
a first order A-N phase transition. At lower values
of $\alpha$ however, the larger range of the nematic phase saturates
the order parameter and the A-N transition becomes
second order. A tricritial point exists at the cross over from second
to first order behavior. The question of the existence of a A-N
tricritical point has motivated interest among researchers for several
years. Since the theoretical discovery by Kobayashi \cite{k1970theory, kobayashi1971theory}
and McMillan \cite{mcmillan1971simple, mcmillan1972x}, there have been several
experiments that strongly suggest such a point does exist on phase
diagrams of temperature vs pressure \cite{mckee1975orientational} or vs concentration of the liquid crystalline material \cite{johnson1975evidence, karat1979elasticity}. Alben \cite{alben1973nematic}
predicted a He$^{3}$-He$^{4}$ like TCP in binary liquid
crystal mixtures. In 1973 Keyes et al \cite{keyes1973tricritical} reported a pressure
study for the transition between the smectic A
and cholesteric liquid crystalline phases of cholesteryl oleyl carbonate
and discovered a new type of TCP. However, Halperin, Lubensky and
Ma \cite{halperin1974first, halperin1974analogy} argued that the A-N transition
can never be truly second order, thus, ruling out conventional tricritical
behavior. The trend of three phases meeting at a single point was
experimentally found by Shashidhar and Chandrasekhar \cite{shashidhar1979high}
when they investigated the pressure dependence for methoxybenzoic
acid and ethoxybenzoic acid. This work not only supported the appearance
of a triple point (TP) in liquid crystalline systems but also established
for the first time the existence of the solid-nematic-isotropic and
solid-smectic A-nematic triple points in a single
component system. This investigation gave a detailed insight on the
effect of pressure on the liquid crystalline phase transitions confirming
the presence of a tricritical point (TCP) at $2.66$ kbar as predicted
by Keyes et al \cite{keyes1973tricritical}. Later in 1979 D. Brisbin \cite{brisbin1979specific}
et al reported the presence of tricritical point in the homologous
series pentylbenzenethioalkoxybenzoate by their specific heat and
birefringence measurements. Thoen \cite{thoen1984nematic} et al in 1984 also
showed by using adiabatic-scanning calorimetry that although A-N
transition is first order for very narrow N ranges, it is second order
for large N ranges in the series alkoxycyanobiphenyl. Up till now
many experiments and theories \cite{marynissen1985experimental, lampe1986high, longa1986tricritical, rananvare1987nematic, huster1987calorimetric, stine1989calorimetric, dasgupta2003splay, mukherjee2002pressure, cladis1981reentrant}
have confirmed this trend of appearances of TP and TCP for many liquid
crystal materials and liquid crystal mixtures.
\paragraph{} In 2018 \cite{dasgupta2018pressure} we had presented a molecular field theory to study pressure induced phase transitions in liquid crystals. A simple effective potential was chosen to discuss in particular the N-I phase transition. In the present paper, we intend to demonstrate the utility of our model in a more general case. Following McMillan's molecular model, we have extended the potential for studying the A$\to$ N$\to$ I phase transitions. The purpose of this paper is to present the pressure dependence of such systems using our model and to calculate its properties. A qualitative as well as quantitative comparison between the results so obtained and those obtained with existing theoretical and experimental results have been made.

\section{Our Model}

One of the pioneering works in theoretical modelling of liquid crystals
was by Maier and Saupe \cite{maier1960simple} who in order to explain the 
orientational order in nematics introduced the potential 

\begin{equation}
U_{MS}=-\frac{A}{v^{2}}\left\langle P_{2}\right\rangle P_{2}(\cos\vartheta)\label{eq:MSpot}
\end{equation}
For any smectic A liquid crystal there is an additional
one-dimensional translational periodicity which requires some degree of translational order in addition to the long-range orientational order for characterizing the phase. McMillan \cite{mcmillan1971simple} developed Eq. \ref{eq:MSpot} further to account for the possibility of a smectic A phase.
 Any realistic theoretical model
of liquid crystals which allow volume fluctuation must include both
repulsive and attractive interactions. In \cite{dasgupta2018pressure}
we had used a molecular field approach by adapting the Maier Saupe
potential for this purpose by adding an isotropic volume dependent
term. In this paper we further extend that work by following the development by McMillan \cite{mcmillan1971simple} that included another term to account for the translational periodicity. We have used this to study the
liquid crystalline smectic A to nematic phase transition.
The effective single particle potential we choose for this purpose
is given by 

\vspace{-0.4cm}

\begin{center}
\begin{equation}
\displaystyle{ 
 U= \frac{u_{4}}{v^{4}} -\frac{u_{2}}{v^{2}}-\frac{Au_{2}}{v^{2}} \left[ \langle P_{2}\right\rangle P_{2}(\cos\vartheta)+ \alpha \left\langle P_{2}(\cos\vartheta)\cos\left(\frac{2\pi z}{d}\right)\right\rangle P_{2}(\cos\theta) \cos \left(\frac{2\pi z}{d}\right)]
}
\end{equation}
\end{center}


\vspace{-0.2cm}
where $u_{4}$, $u_{2}$, $A$ and $\alpha$ are constants and $v$
is the volume of the fluid per molecule. Here the volume dependence
of the isotropic terms has been chosen to mimic the scaling behavior
of the familiar Lennard-Jones potential \cite{dasgupta2018pressure}. Both Maier Saupe and McMillan models were on fixed lattices. In the current paper we investigate a continuum version. We have considered the isothermal-isobaric ensemble, denoted as the NPT ensemble, here.

Using this potential the canonical partition function
can be written as

\begin{equation}
\displaystyle{ Z_{NVT}=\frac{V^{N}}{N!\Lambda^{3N}}\left(\overline{Z}_{1}\exp\left(\frac{\beta}{2}\left\langle U\right\rangle \right)\right)^{N}\exp\left[-N\beta\left(\frac{u_{4}}{v^{4}}-\frac{u_{2}}{v^{2}}\right)\right]}
\label{eq:ZNVT}
\end{equation}
\vspace{0.5cm}

where 

\begin{equation}
\overline{Z}_{1} = \frac{1}{d}\int_{0}^{d}\int_{0}^{1} I \left(\cos\vartheta, z\right) d\left(\cos\vartheta \right) dz \label{eq:single particle part function}
\end{equation}
and
\begin{equation}
I \left(\cos\vartheta, z\right)= \exp \left( 
\frac{\beta Au_{2}}{v^{2}}
\left( \langle P_{2} \rangle P_{2} \left( \cos\vartheta \right) + \alpha \langle P_{2}(\cos\vartheta)\cos\left(\frac{2\pi z}{d} \right)\rangle \left( P_{2} (\cos\vartheta)\cos\left(\frac{2\pi z}{d} \right) \right) \right)
 \right). \nonumber
\end{equation}

Here, $\beta=1/KT$, $d$ is the layer thickness and $V$ is the volume
of the system.

Hence, the partition function for the NPT ensemble is
\begin{equation}
Z_{NPT}=\int_{0}^{\infty}\frac{1}{V_{0}}\exp(-\beta pV)Z_{NVT\,}dV\label{eq:ZNPT}
\end{equation}
which becomes
\begin{equation}
Z_{NPT}=\frac{N^{N+1}}{N!V_{0}\Lambda^{3N}}\int_{0}^{\infty}\exp\left(Nf(v)\right)dv\label{eq:partfunc1}
\end{equation}
where 
\begin{equation}
f(v)=-\beta pv+\ln v+\ln\overline{Z}_{1}-\frac{Au_{2}\beta}{2v^{2}}\biggl(\eta^{2}+\alpha\sigma^{2}\biggr)-\frac{\beta}{2}\biggl(\frac{u_{4}}{v^{4}}-\frac{u_{2}}{v^{2}}\biggr)\label{eq:8f(v)}
\end{equation}
where $\eta=\left\langle P_{2}(\cos\vartheta)\right\rangle $ denotes
the purely orientational (i.e. nematic) order parameter and $\sigma=\left\langle P_{2}(\cos\vartheta)\cos\left(\frac{2\pi z}{d}\right)\right\rangle $ denotes
the mixed order parameter which describes the coupling between the
degrees of orientational and translational order. As in McMillan \cite{mcmillan1971simple}
we neglect the purely translational order parameter. 

So $f(v)$ depends on $v$, $\left\langle P_{2}\left(\cos\vartheta\right)\right\rangle $, 
$\left\langle P_{2}\left(\cos\vartheta\right)\cos\left(\frac{2\pi z}{d}\right)\right\rangle $
and on the constant parameters $p$, $\beta$, $u_{2}$, $u_{4}$, $\alpha$ and $A$. Here we are singling out its volume dependence since we are
going to use the saddle point method where the integral is only over
volume. Since $N\gg1$ we can use the saddle point approximation \cite{dasgupta2018pressure}
to write (upto a multiplicative constant which we ignore)
\begin{equation}
Z_{NPT}=\frac{N^{N+1}}{N!V_{0}\Lambda^{3N}}\exp(Nf(v_{*}))\label{eq:partfunc2}
\end{equation}
where $v_{*}$ is the value of $v$ which maximizes $f(v)$. 

Maximizing $f(v)$ we get the equation of state
\begin{equation}
p-\frac{1}{\beta v_{*}}-\frac{2u_{4}}{v_{*}^{5}}+\frac{u_{2}}{v_{*}^{3}}\left[1+A\eta^{2}+A\alpha\sigma^{2}\right]=0\label{eq:differentiatingfree energy}
\end{equation}
Using this $f(v_{*})$ we can construct the Gibb's free energy of
the system given by
\[
G\beta=-Nf(v_{*})-\frac{3N}{2}\ln\beta
\]
up to a constant. Hence the Gibb's free energy of the system in units
of $KT$ is given by
\begin{equation}
G\beta=N\left(p\beta v_{*}-\ln v_{*}+\frac{\beta}{2}\frac{u_{4}}{v_{*}^{4}}-\frac{\beta u_{2}}{2v_{*}^{2}}-\ln\overline{Z}_{1}\left(\lambda,\mu\right)+\frac{\beta}{2}\left(\frac{Au_{2}\eta^{2}}{v_{*}^{2}}+\frac{Au_{2}\alpha\sigma^{2}}{v_{*}^{2}}\right)\right)-\frac{3N}{2}\ln\beta\label{eq:gibb's free energy}
\end{equation}

where $\lambda=\frac{Au_{2}\beta}{v^{2}}\eta$ and $\mu=\frac{Au_{2}\beta\alpha}{v^{2}}\sigma$.

Now by minimizing Eq. \ref{eq:gibb's free energy}
w.r.t $\lambda$ and $\mu$ we obtain 
\begin{equation}
\eta=\frac{1}{\overline{Z}_{1}}\left(\frac{\partial\overline{Z}_{1}(\lambda,\mu)}{\partial\lambda}\right)\label{eq:nematicorder parm}
\end{equation}
and
\begin{equation}
\sigma=\frac{1}{\overline{Z}_{1}}\left(\frac{\partial\overline{Z}_{1}}{\partial\mu}\right)\label{eq:crossed order parm}
\end{equation}
Solving Eq. \ref{eq:differentiatingfree energy}, Eq. \ref{eq:nematicorder parm}
and Eq. \ref{eq:crossed order parm} simultaneously we obtain the value
of $\eta$, $\sigma$ and $\nu_{*}$ for the different liquid crystalline
phases. Again solving Eq. \ref{eq:differentiatingfree energy} for
$\eta=\sigma=0$ yields the value of $\nu_{*}$ for the isotropic
phase. Similarly the condition $\eta=0$ and $\sigma\ne0$ yields
the smectic A phase and $\sigma=0$ and $\eta\ne0$
yields the nematic phase. Out of these possibilities the equilibrium
value of $\eta$, $\sigma$ and $\nu_{*}$ for a particular set of
constant parameters $p$, $\beta$, $u_{2}$, $u_{4}$, $\alpha$
and $A$ is decided by checking which of these yield the lower value
of $G$. This value of $v_{*}$, $\eta$ and $\sigma$ globally minimizes
the function $G$. 

Now we obtain the critical constants $p_{c}$, $\beta_{c}$
and $\nu_{c}$ for the isotropic-vapor transition from Eq. \ref{eq:differentiatingfree energy}
(setting $\eta=\sigma=0$) in terms of $u_{2}$ and $u_{4}$. 
\begin{equation}
\nu_{c}^{2}=\frac{20u_{4}}{3u_{2}}\label{eq:critconst2}
\end{equation}
\begin{equation}
\frac{1}{\beta_{c}}=\frac{9u_{2}^{2}}{40u_{4}}\label{eq:critconst3}
\end{equation}
\begin{equation}
p_{c}\beta_{c}\nu_{c}=\frac{8}{15}\label{eq:critconst4}
\end{equation}

To get rid of the constants $u_{2}$ and $u_{4}$, we have expressed the temperature ($T$), pressure ($p$) and volume ($\nu$) in terms of reduced parameters 
$\vartheta=\frac{T}{T_{c}}$, 
$\pi=\frac{p}{p_{c}}$, 
$\omega=\frac{\nu}{\nu_{c}}$. 

From Eq. \ref{eq:differentiatingfree energy} we
obtain the reduced equation of state
\begin{equation}
\pi=\frac{15\vartheta}{8\omega}+\frac{3}{8\omega^{5}}-\frac{5}{4\omega^{3}}\left(1+A\eta^{2}+A\alpha\sigma^{2}\right)\label{eq:reduced eq of state}
\end{equation}
The corresponding free energy in terms of the reduced set of parameters
$\pi$, $\omega$ and $\vartheta$ is 
\begin{equation}
G_{R}=\frac{8\pi\omega}{15\vartheta}-\ln\omega+\frac{1}{20\vartheta\omega^{4}}+\frac{1}{3\vartheta\omega^{2}}\left(A\eta^{2}+A\alpha\sigma^{2}-1\right)-\ln\overline{Z}_{1}\left(\lambda,\mu\right)+\frac{3}{2}\ln\vartheta\label{eq:free reduced}
\end{equation}
Using Eq. \ref{eq:reduced eq of state} and Eq. \ref{eq:free reduced},
we obtained the values of $\eta$, $\sigma$ and $\omega$. Again
solving Eq. \ref{eq:reduced eq of state} for $\eta=\sigma=0$ yields
the value of $\omega$ for the isotropic phase. As discussed earlier
out of these two values the equilibrium value of $\eta$, $\sigma$
and $\omega$ is decided on the basis of which of these values yields
a lower $G_{R}$. 

The entropy in the reduced form is
\begin{equation}
S=\frac{S^{'}}{Nk}=-\frac{2A}{3\vartheta\nu^{2}}\left(\eta^{2}+\alpha\sigma^{2}\right)+\ln\left(\overline{Z}_{1}\left(\lambda,\mu\right)\right)\label{eq:entropy}
\end{equation}

We have calculated the specific heat at constant
pressure using the equation
\begin{equation}
C_{p}=T\left(\frac{\partial S}{\partial T}\right)_{p}\label{eq:specificheat}
\end{equation}
where the derivative has been evaluated numerically.

\section{Results and Discussions}

In this section, the results obtained using our model potential have been discussed and compared with the existing theoretical and experimental results. In our model, $A$ and $\alpha$ are free parameters which can be
varied to match different liquid crystalline systems. For example
the value of $A=0.675$ and $\alpha=0.494$  helped us fit the experimental
data of A-N transition temperature
($T_{AN}=343$K) and the N-I transition temperature
($T_{NI}=353$K) for 8OCB quite well. 

\subsection{Effect of pressure on the order parameters:}

\begin{figure}
\begin{centering}
\includegraphics[width=0.45\columnwidth]{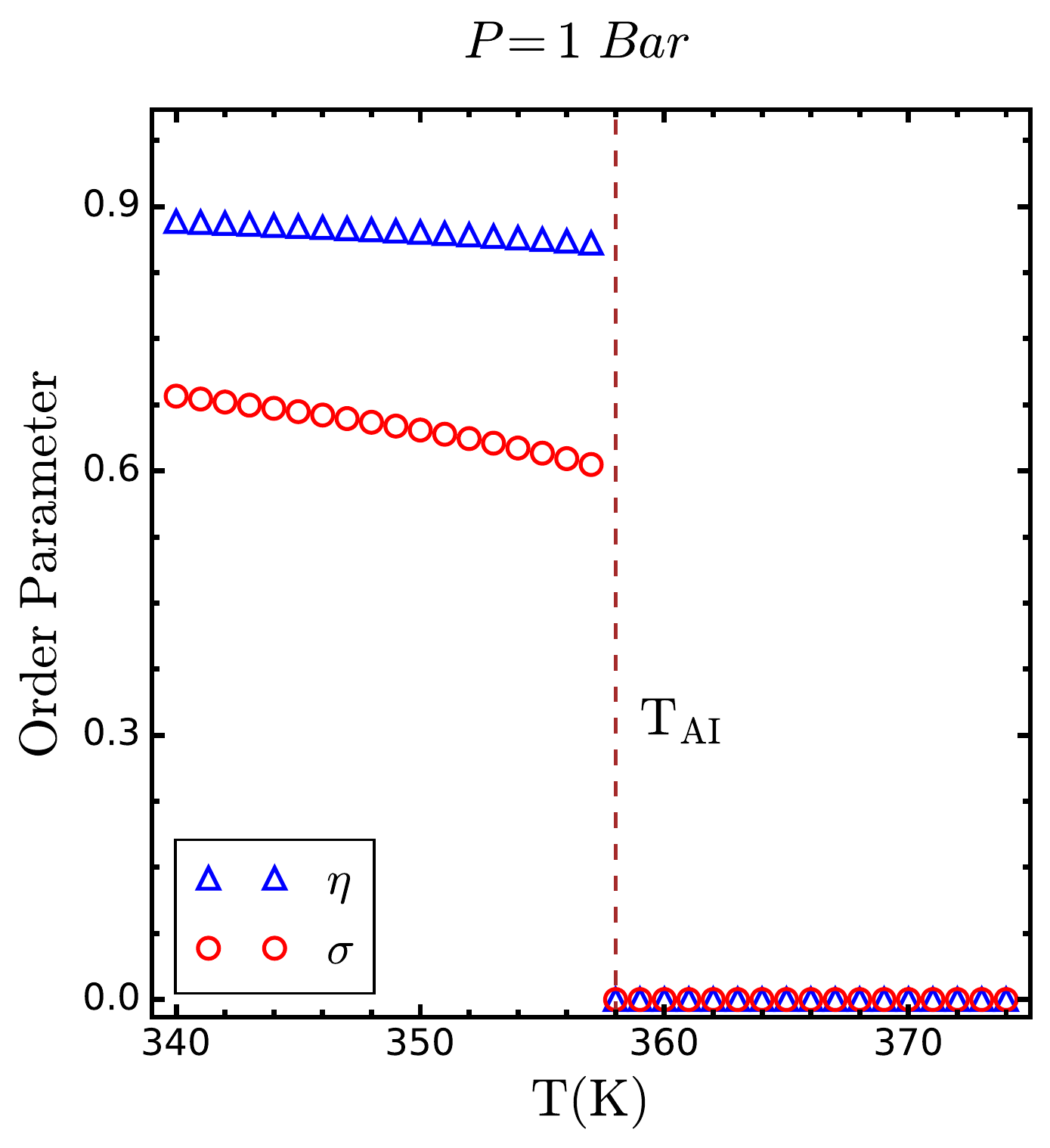}\includegraphics[width=0.45\columnwidth]{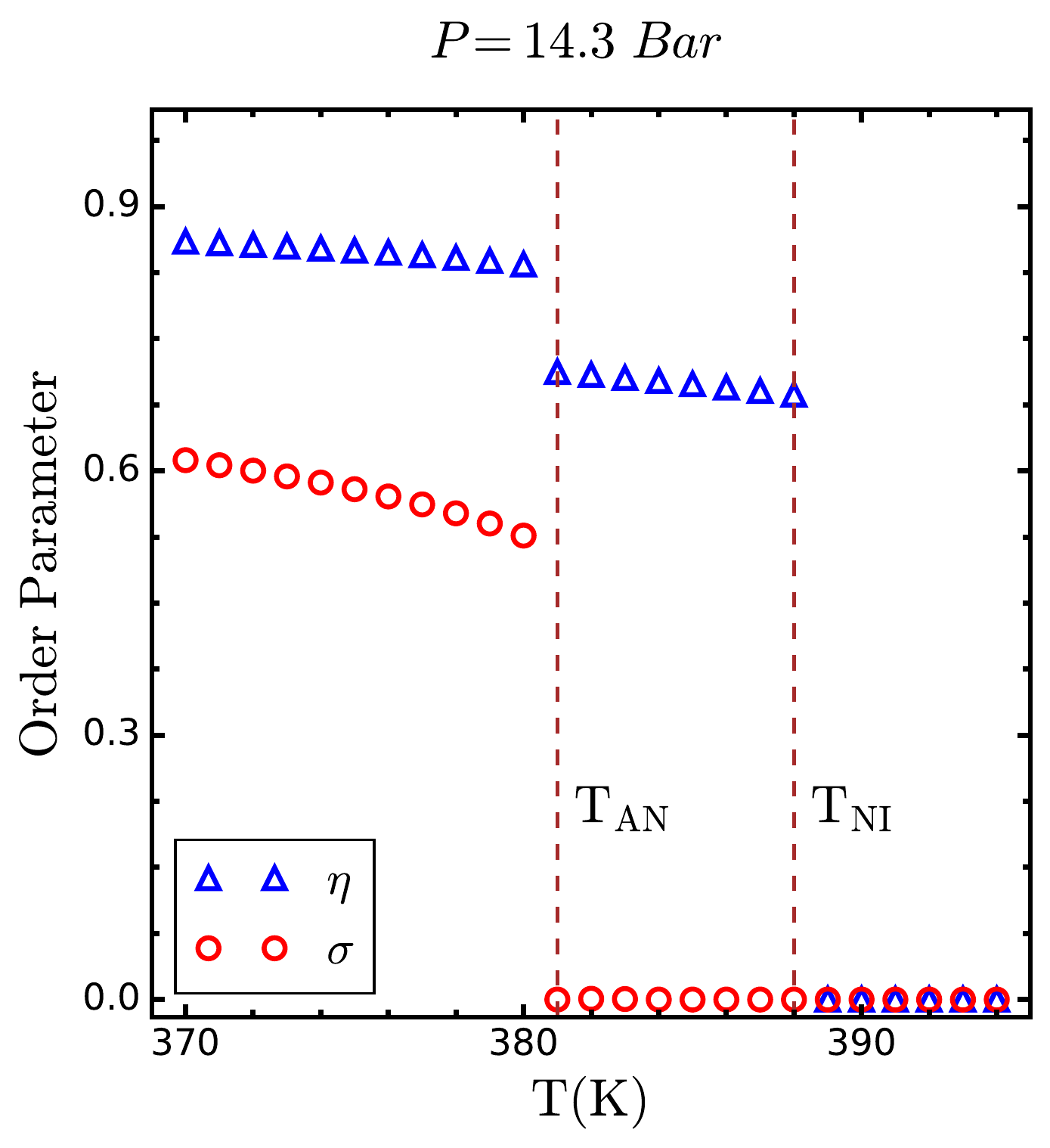}
\par\end{centering}

\caption{\label{fig:effectOFpressure}Variation of order parameters $\eta$ and $\sigma$ with temperature at different values of pressure.}
\end{figure}

The variation of order parameters $\eta$ and $\sigma$ with temperature are shown in fig. \ref{fig:effectOFpressure} for the choice of model parameter $\alpha=0.55$. Variation at two different pressure has been discussed here. The left diagram shows variation at $P=1 $ Bar and the right one at $P=14.3 $ Bar. The effect of pressure on the A-I or A-N transition is to increase the transition temperature as can be seen from the figure. This is because of the fact that an increase in pressure brings about more order in the liquid crystal molecules. From the variation of the $\sigma$ curve, the A-I or A-N transition is found to be of first order even at elevated pressure. It is also seen that as pressure increases, the discontinuity of the $\sigma$ curve, at the transition, decreases. This result is in accordance with the work by Mukherjee et al \cite{mukherjee2002pressure} in 2002. Another interesting feature is the appearence of the pressure induced nematic phase at higher pressure as can be seen following the $\eta$ curve.  

\subsection{Effect of temperature on the order parameters}

\begin{figure}
\begin{centering}
\includegraphics[height=18\baselineskip]{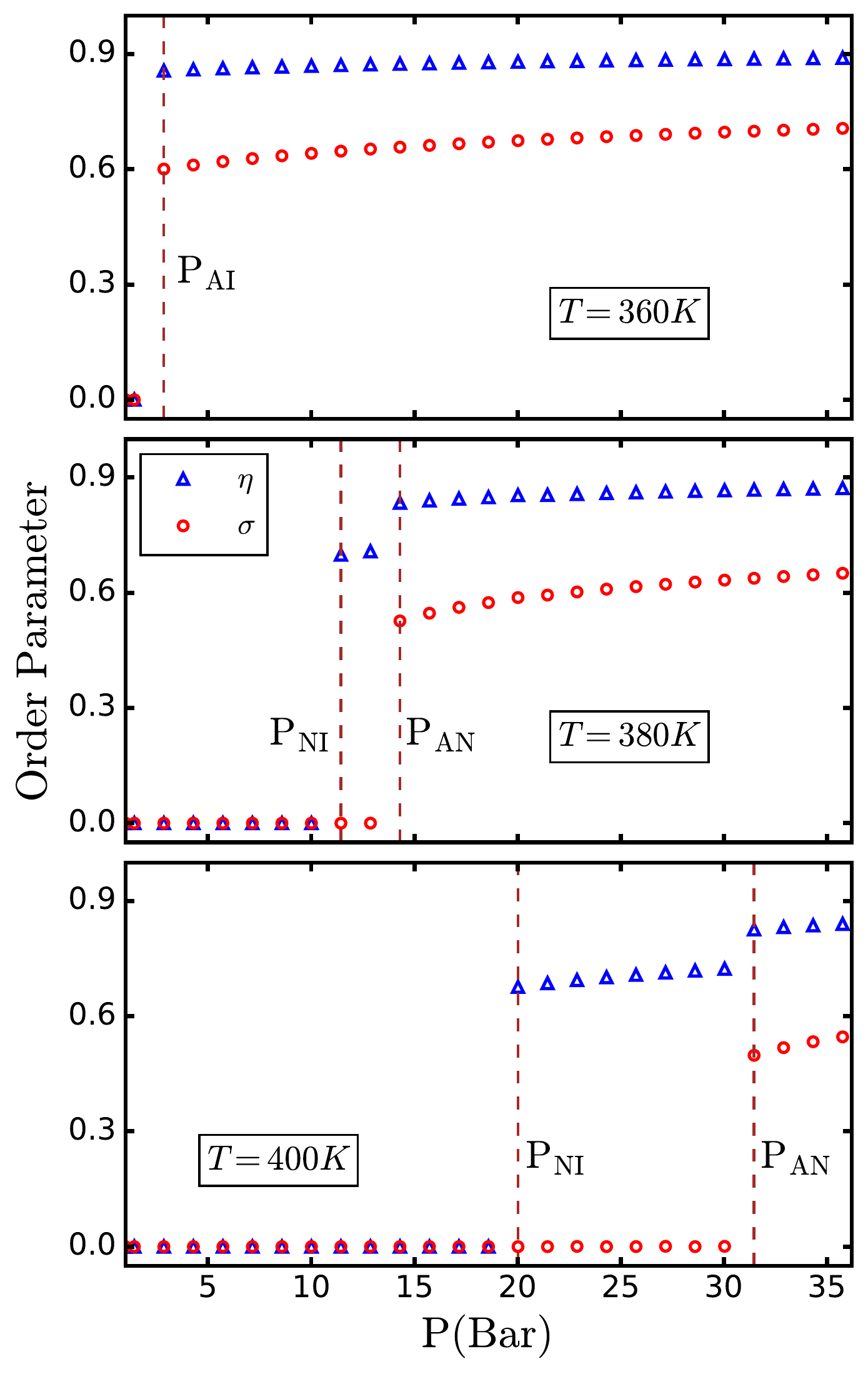}
\par\end{centering}
{\caption{\label{fig:effectOFtemp}Variation of order parameters $\eta$ and $\sigma$
with pressure at various fixed values of temperature.}
}
\end{figure}

The pressure dependence of the order parameters $\eta$ and $\sigma$ is shown in fig. \ref{fig:effectOFtemp} for three representative values of temperature.  At $T=360K$ there is a direct smectic A to isotropic transition while at the higher T values i.e. at $T=380$K and $T=400$K nematic phase is also present. Here, as expected, we can see the A-N transition at $P_{AN}$ and subsequently a N-I transition at $P_{NI}$ for the lower values of pressure.
It can be noted that the value of order parameters decreases with the decrease of pressure in each case thereby showing that the liquid crystalline system becomes less and less ordered as pressure decreases.

\subsection{Significance of the model parameter $\alpha$}

The physical parameter
$\alpha$ acts as a dimensionless interaction
strength for the smectic A phase. From the theoretical point of view $\alpha$ is related to the length of the alkyl chain in a homologous series such that $\alpha$ should increase with increasing chain length. In analogy to the McMillan theory \cite{mcmillan1971simple} we have taken different values of the constant $\alpha$ to discuss the variation of liquid-crystal behavior in a homologous series.

\subsubsection{$\alpha$ Phase diagram}

\begin{figure}
\begin{centering}
\includegraphics[height=18\baselineskip]{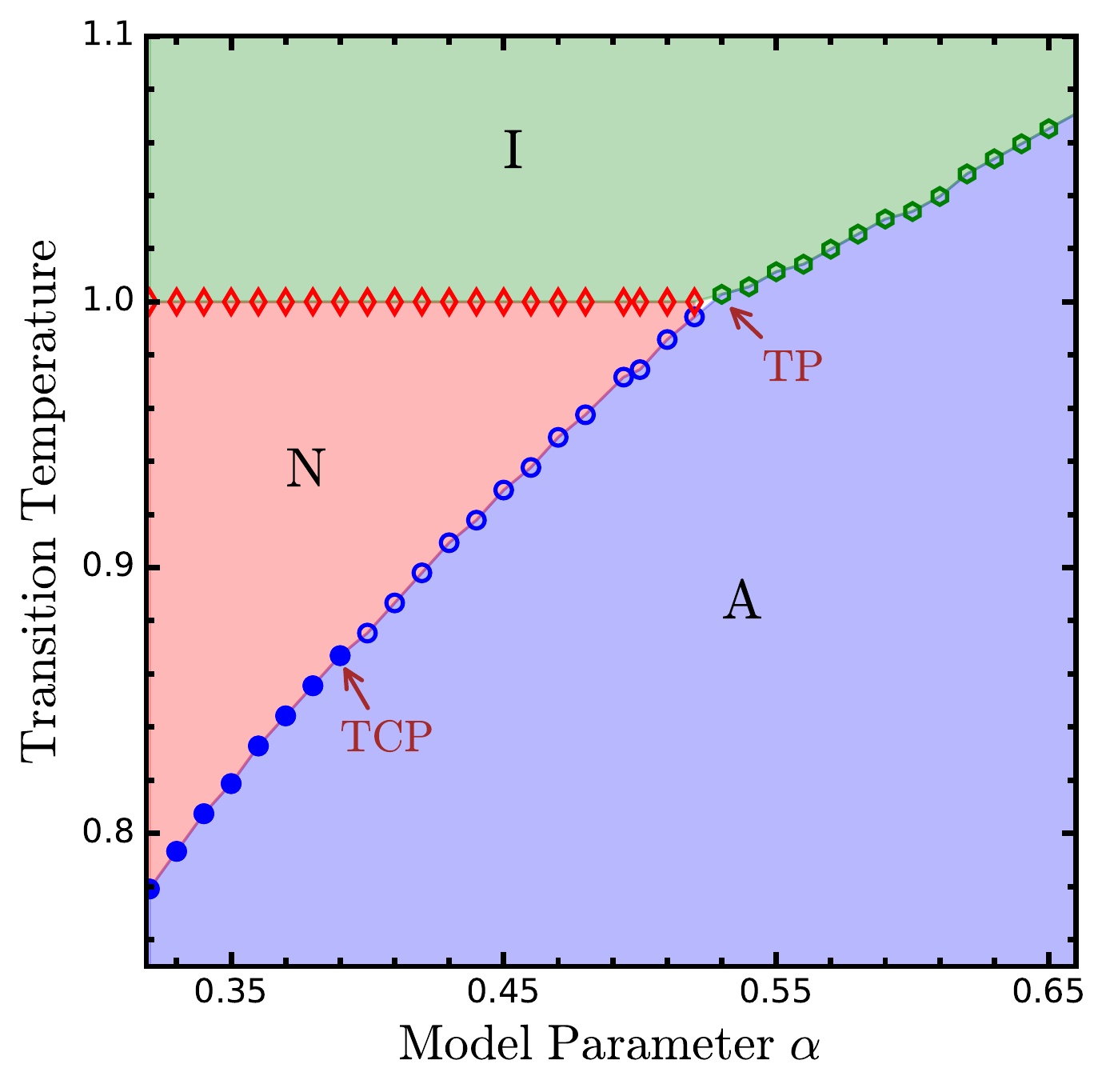}
\par\end{centering}
\caption{\label{fig:Phasediagram_alpha}Phase diagram with the variation of the model parameter $\alpha$ at atmospheric pressure. The lines seperate the regions of stability of the smectic A (A), nematic (N) and isotropic (I) phases. Red line corresponds to the N-I transition, the blue line is the A-N transition and the green line is the A-I transition. Filled and open blue circles on the A-N transition line indicate second order and first order phase transitions respectively. }
\end{figure}

The transition temperatures ($\frac{T}{T_{NI}}$) as a function of the model parameter $\alpha$ are plotted in the fig. \ref{fig:Phasediagram_alpha}, where $T_{NI}=353K$, the N-I transition temperature. The phase diagram shows that at lower values of $\alpha$ (i.e. for shorter chain length) all three
phases, namely, smectic A, nematic and isotropic exist and,
A$\to$ N$\to$ I phase transitions occur as temperature is increased. For higher values of $\alpha$ ($\alpha>0.53$) (i.e. for longer chain length), only smectic A and isotropic phases can be observed
and a direct A$\to$ I transition takes place.
$\frac{T}{T_{NI}}=1$ represents the N-I transition line upto $\alpha=0.53$. The A-N transition temperature is an increasing function of $\alpha$ and reaches the N-I line at $\alpha=0.53$.
 All these three phase transition lines, A-N, N-I and A-I, as can be seen from the fig. \ref{fig:Phasediagram_alpha}., 
meet at ($\alpha=0.53$, $\frac{T}{T_{NI}}=1$), and thus, form a triple point (TP). The $T_{AN}$ curve is very nearly a continuation of the $T_{AI}$ curve at the triple point as expected from the theoretical model.
 
Another interesting feature can be pointed out from this phase diagram. The A-N transition line is second order for $\alpha<0.4$ (this has been indicated by the filled blue circles in the phase diagram) and
is first order for $0.4<\alpha<0.53$ (this has been indicated by the open blue circles in the phase diagram). This confirms the presence of a tricritical point (TCP) at $\alpha=0.39$ and $\frac{T_{AN}}{T_{NI}}>0.866$. This phase diagram is seen to give an excellent qualitative agreement with McMillan's model \cite{mcmillan1971simple}.

\subsubsection{ Thermodynamic variables at different values of $\alpha$ :}

To illustrate the above stated behavior of $\alpha$ phase diagram, in this section we shall discuss the order parameters $\eta$ and $\sigma$, the entropy $S$ and the specific heat $C_{P}$ as a function of temperature for  three different values of interaction strength $\alpha$.

\begin{figure}
\begin{centering}
\includegraphics[height=15\baselineskip]{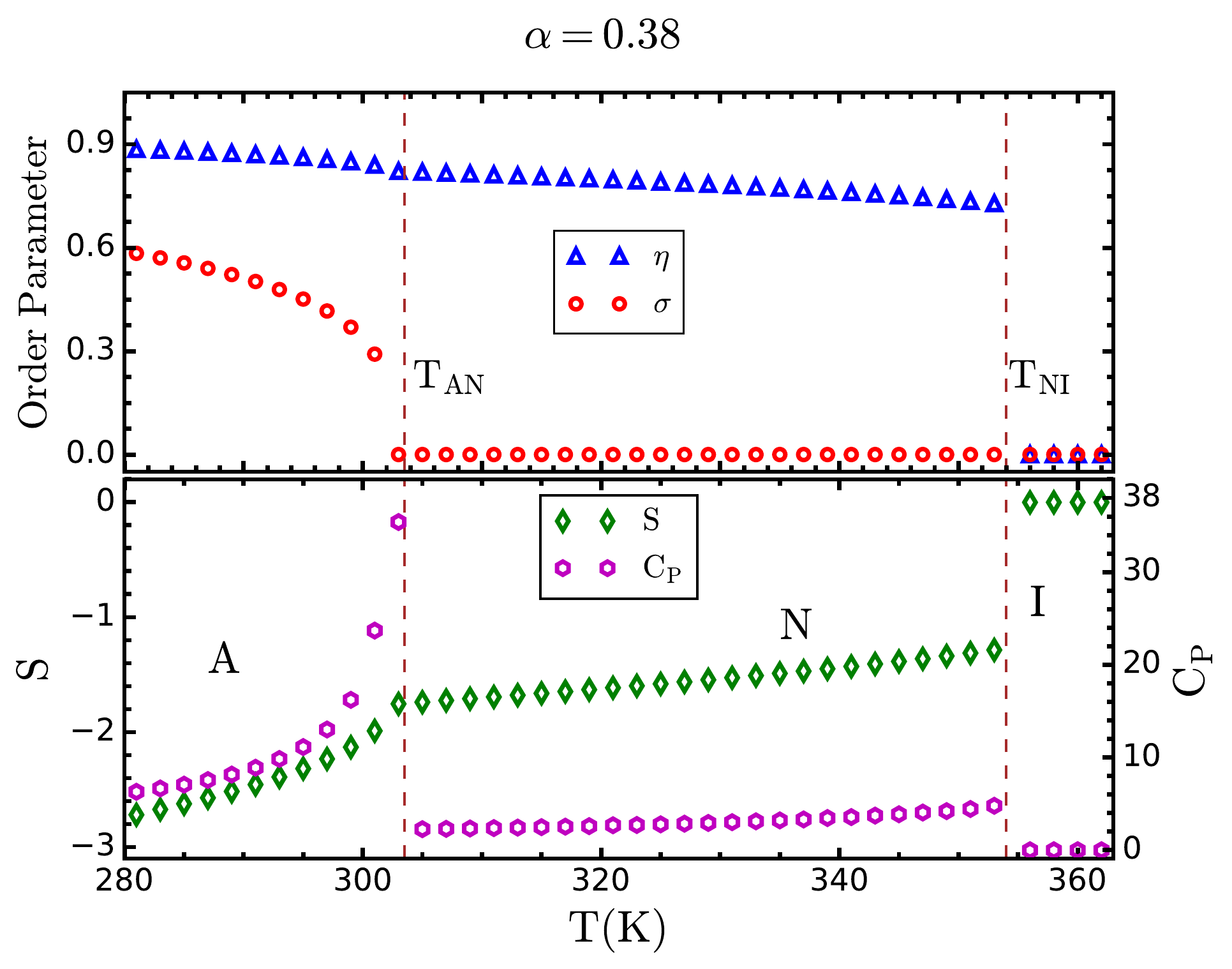}
\par\end{centering}
\textcolor{black}{\caption{\label{fig:orderparm-T_alpha_0.38}Variation of order parameters $\sigma$, $\eta$, entropy $S$ and specific heat $C_{P}$ with temperature for $\alpha=0.38$ showing the second order smectic-A-nematic transition and the first order nematic-isotropic liquid transition at the atmospheric pressure. }
}
\end{figure}
\begin{figure}
\begin{centering}
\includegraphics[height=15\baselineskip]{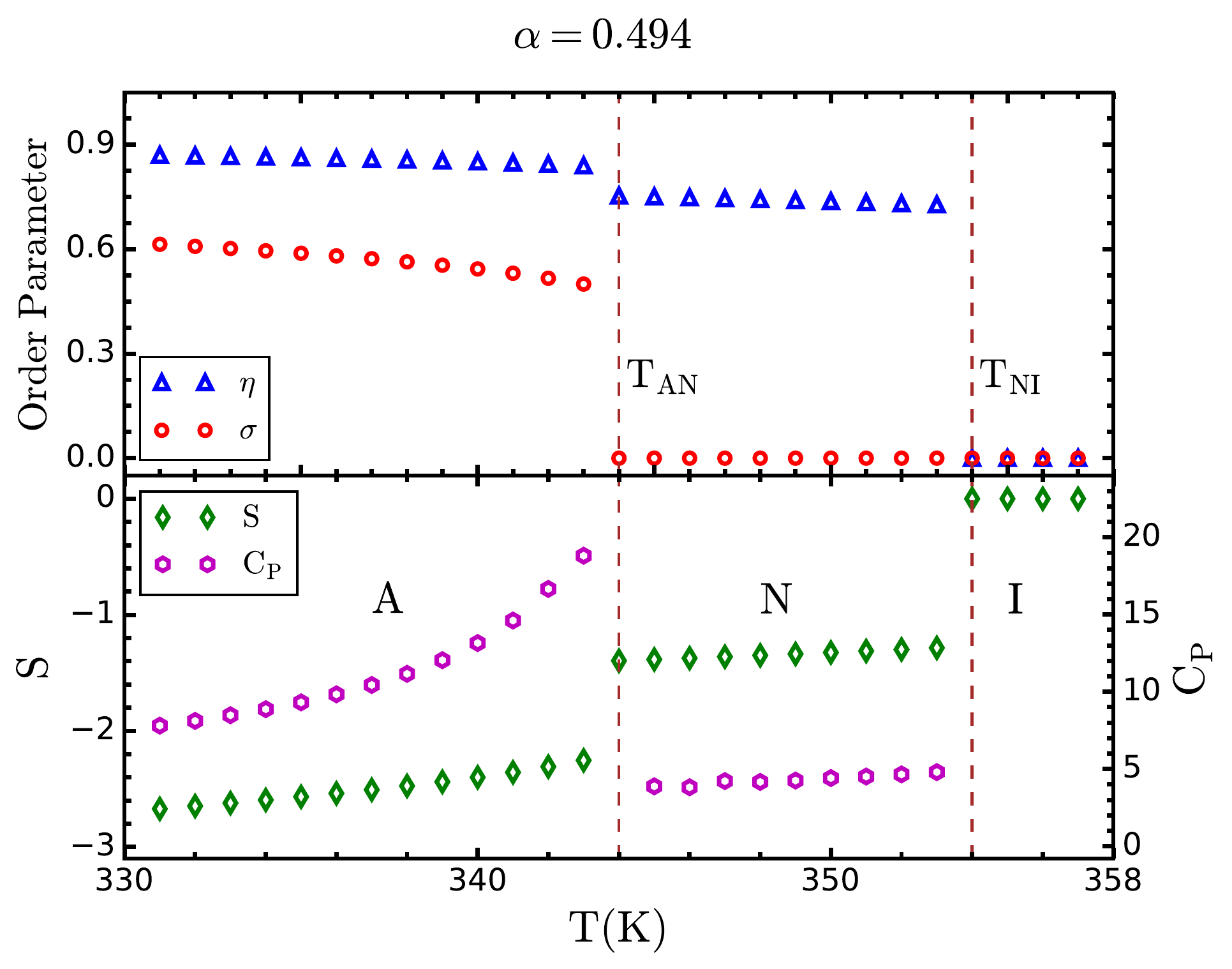}
\par\end{centering}
\textcolor{black}{\caption{\label{fig:orderparm-T_alpha_0.494}Variation of order parameters $\sigma$, $\eta$, entropy $S$ and specific heat $C_{P}$ with temperature for $\alpha=0.494$ showing the first order smectic-A-nematic transition at the atmospheric pressure.}
}
\end{figure}

For $\alpha=0.38$ (fig. \ref{fig:orderparm-T_alpha_0.38}), the A-N transition is second order as the smectic order parameter $\sigma$ falls continuously to $0$ at the transition ($\frac{T_{AN}}{T_{NI}}=0.85$). The corresponding N-I transition is, however, first order in nature because of the presence of a discontinuous jump at $T_{NI}$. Entropy ($S$) and specific heat ($C_{P}$) also show similar kind of behavior. We can see that the entropy changes continuously at the A-N transition indicating a second order but has a discontinuous jump at the N-I transition favoring first order phase transition. In case of $C_{P}$, we get discontinuous jump at both the transitions but quantitatively the value of discontinuity at the N-I transition is larger than that at the A-N transition. 

For $\alpha=0.494$ (fig. \ref{fig:orderparm-T_alpha_0.494}) the A-N transition is first order as $\sigma$ drops discontinuously to $0$ at the transition ($\frac{T_{AN}}{T_{NI}}=0.97$). The corresponding N-I transition is also first order in nature. The discontinuity in $S$ and $C_{P}$ at both transitions show that they are first order transitions.

\begin{figure}
\begin{centering}
\includegraphics[height=15\baselineskip]{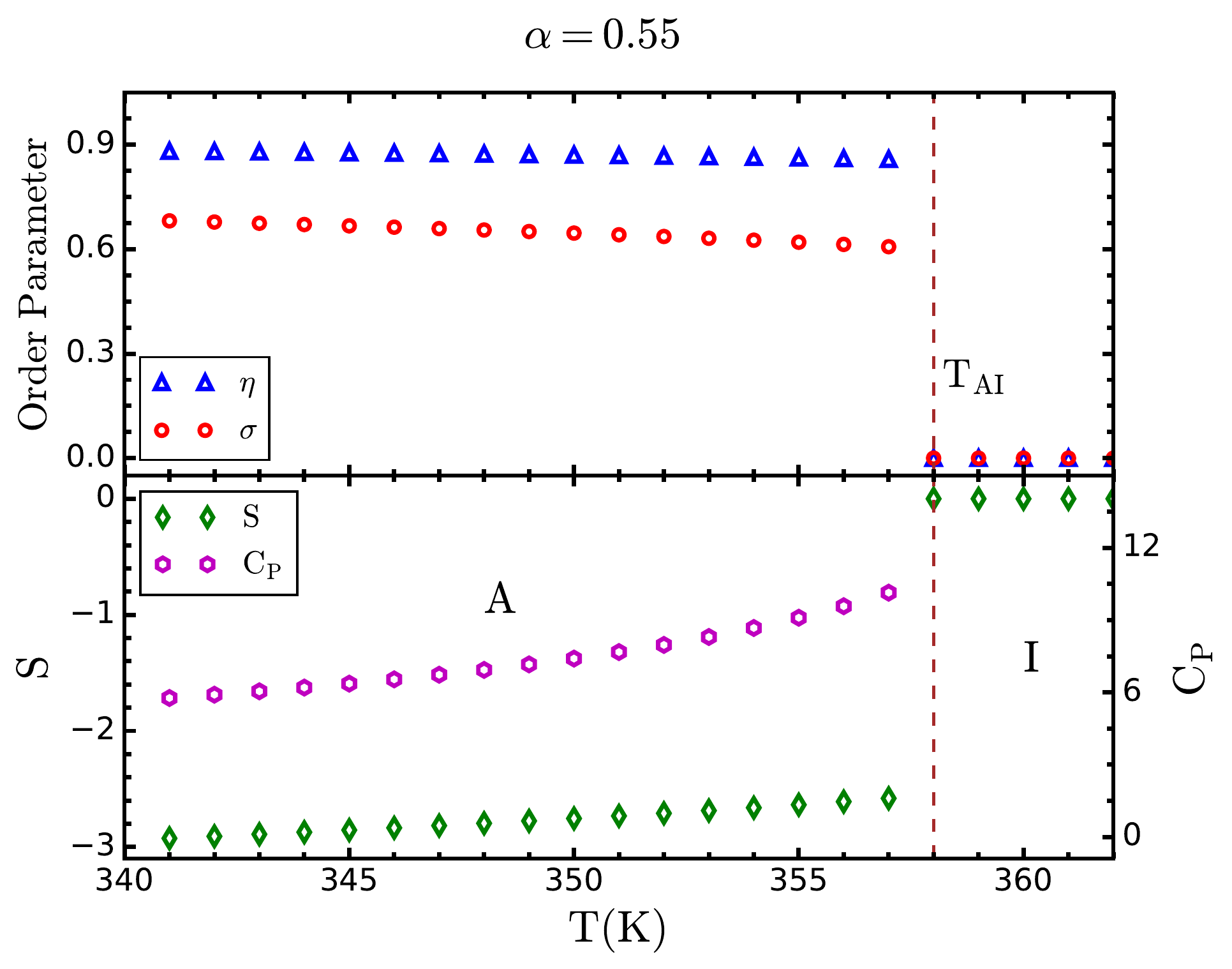}
\par\end{centering}
\textcolor{black}{\caption{\label{fig:orderparm-T_alpha_0.55}Variation of order parameters $\sigma$, $\eta$, entropy $S$ and specific heat $C_{P}$ with temperature for $\alpha=0.55$ showing the first order smectic-A-isotropic liquid transition at the atmospheric pressure.}
}
\end{figure}

For $\alpha=0.55$ (fig. \ref{fig:orderparm-T_alpha_0.55}) a direct A-I transition occurs. Both order parameters $\sigma$ and $\eta$ drops discontinuously to $0$ at $T_{AI}$. The discontinuity in $S$ and $C_{P}$ show that the A-I transition is a first order transition.

All these values are at the atmospheric pressure $P=1$ Bar.

\subsubsection{ Phase diagrams at different values of $\alpha$ :}

A detailed investation on the pressure and temperature
dependence of a liquid crystalline system at different values of $\alpha$
has been carried out. The study shows that smectic A-nematic and nematic-isotropic liquid transition temperatures depend
on pressure as well as on the parameter value $\alpha$. As $\alpha$
value changes the nature of phase diagram changes drastically. Here
we have discussed the phase diagrams for three different values of
$\alpha$. 

\begin{figure}
\begin{centering}
\includegraphics[height=15\baselineskip]{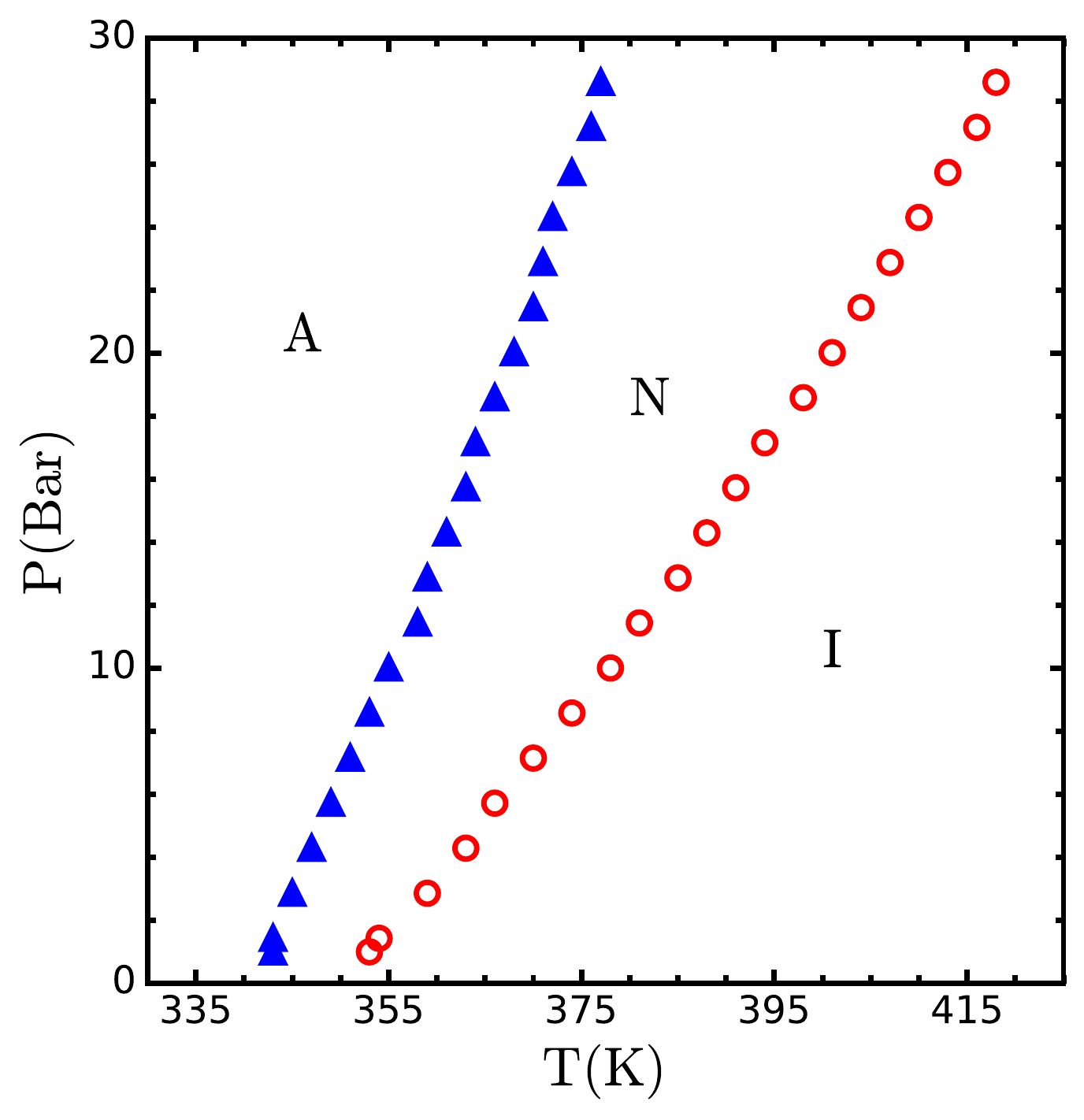}
\par\end{centering}
\textcolor{black}{\caption{\label{fig:Phase-diagram_alpha0.494}Phase diagram for the model parameter $\alpha=0.494$ showing the smectic A, nematic and isotropic
liquid phases. }
}
\end{figure}

\begin{figure}
\begin{centering}
\includegraphics[height=15\baselineskip]{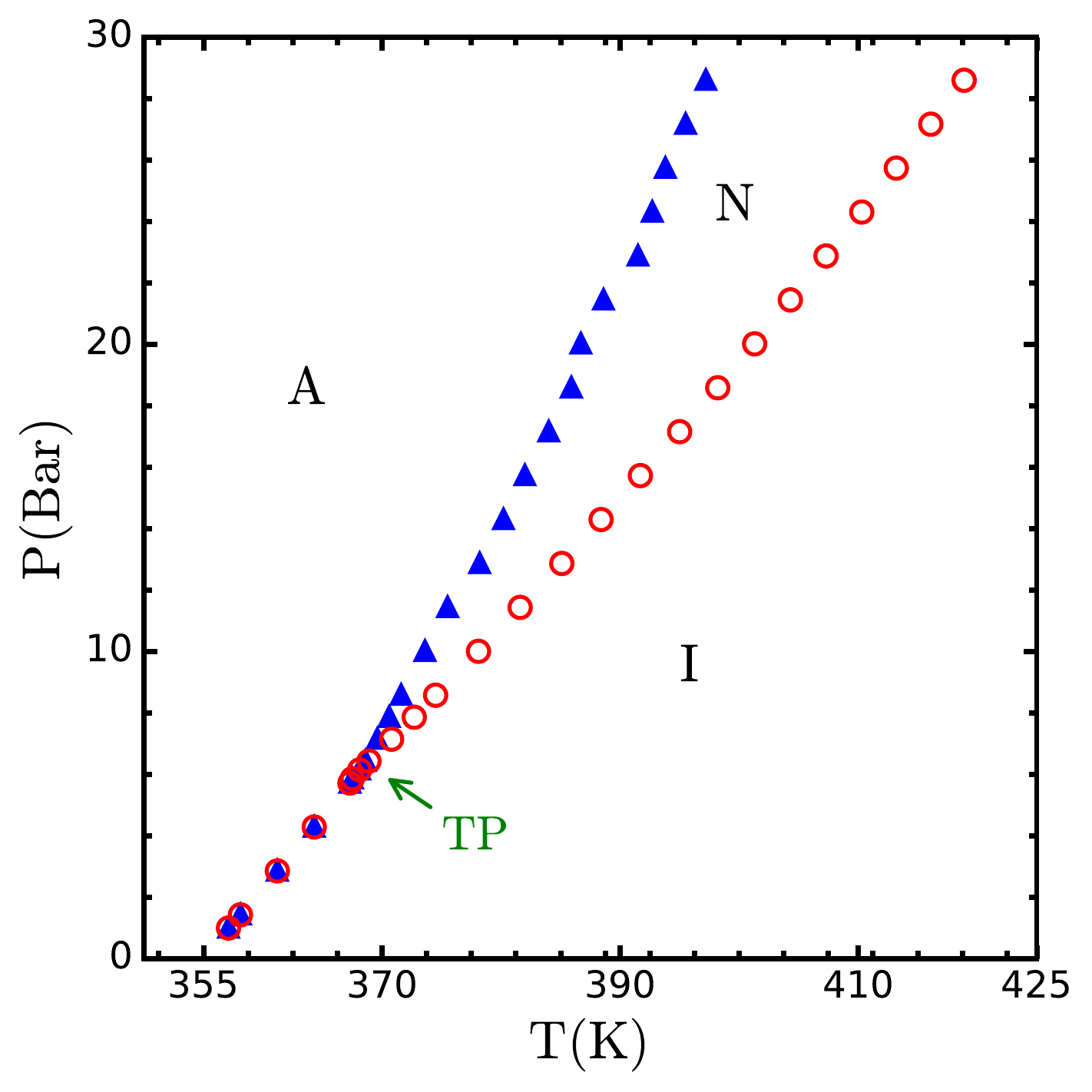}
\par\end{centering}
\textcolor{black}{\caption{\label{fig:Phase-diagram_alpha0.55}Phase diagram for the model parameter $\alpha=0.55$. The intersection of the phase boundaries showing the smectic A-nematic-isotropic triple point (TP) at $368.12$ K and $6.14$ Bar.}
}
\end{figure}

\begin{figure}
\begin{centering}
\includegraphics[height=15\baselineskip]{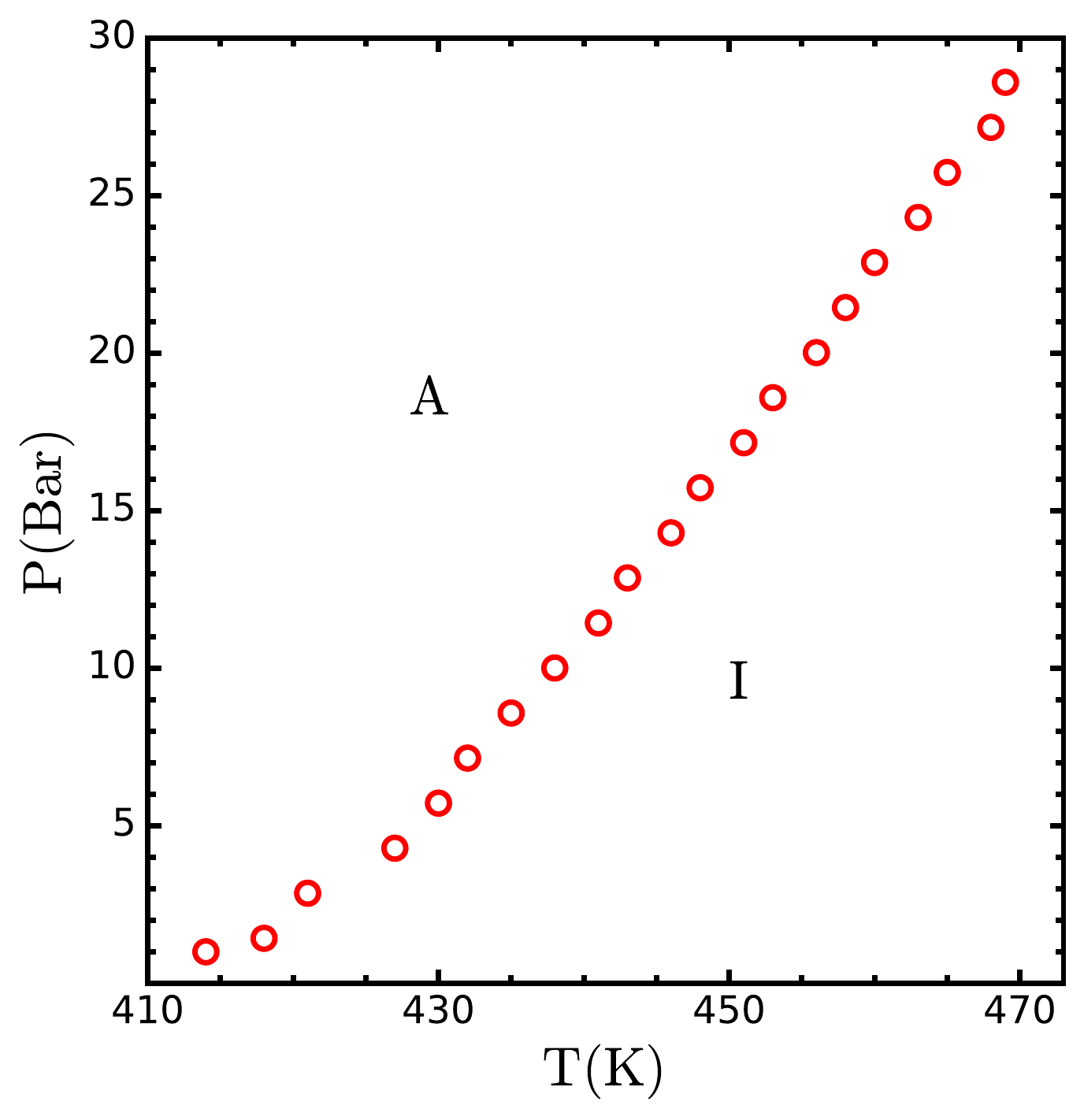}
\par\end{centering}
\textcolor{black}{\caption{\label{fig:Phase-diagram_alpha0.85}Phase diagram for the model parameter $\alpha=0.85$ showing the
smectic A and isotropic liquid phases.}
}
\end{figure}
For $\alpha=0.494$ (fig. \ref{fig:Phase-diagram_alpha0.494}), we have the phase diagram denoting the smectic A, nematic and isotropic liquid phases. As the pressure is raised, both the A-N and N-I transition temperatures increase. It is also seen from the diagram that the slope $\frac{dT}{dP}$ for the N-I transition line is greater than that of A-N transition line. This result is in accordance with the experimental data. This kind of behavior is also expected according to the Clausius-Clapeyron equation. For our choice of parameters, this phase diagram (fig. \ref{fig:Phase-diagram_alpha0.494}) reproduces the known behavior of cyano-octyloxybiphenyl ($8OCB$) as shown by Cladis et. al. \cite{cladis1981reentrant} in their experimental work.

The phase diagram for the model parameter $\alpha=0.55$ is shown in
(fig. \ref{fig:Phase-diagram_alpha0.55}). At lower pressure, there is only one transition, namely, smectic A to isotropic (A-I) transition. At higher pressure the nematic phase appears and there are two transitions, namely, A-N and N-I transitions. The branching point from where A-N and N-I transition lines originate from the A-I transition line is called the triple point (TP). With our specific choice of parameter values, we get TP at ($368.12$ K, $6.14$ Bar). Experimentally, the appearence of such smectic A-nematic-isotropic TP  was found by Lampe et. al. \cite{lampe1986high} in the ninth members of the homologous series of di-alkylazoxybenzenes ($9AB$). 

With higher value of $\alpha$ the nature of phase diagram changes completely. Fig. \ref{fig:Phase-diagram_alpha0.85} shows
the phase diagram for $\alpha=0.85$. Here we can see that our system undergoes a direct transition from the smectic A to the isotropic
phase without going through the nematic phase. Qualitatively, this can be attributed to the high value of $\alpha$ which in turn signifies liquid crystalline molecules with long alkyl chains. This can be seen in the homologous series of cyanobiphenyls (for example $10CB$ or $12CB$ as shown in \cite{drozd2000quasicritical}).

\section{Conclusions}
We have presented a simple model potential which reproduces much of the known behavior of a smectic A-nematic-isotropic liquid phase transition. Investigating the properties of thermotropic liquid crystalline system we have shown that the application of pressure can result in the appearance of a nematic phase. Detailed investigation at different values of $\alpha$ helped us to confirm that large  $\alpha$ value corresponds to long alkyl chain behavior in the homologous series. Different values of the parameters A and $\alpha$ can be chosen to study different liquid crystalline materials.

\section{Acknowledgements}

Sabana Shabnam acknowledges financial support from the Department
of Atomic Energy, Government of India and NISER HPC facility. The
author also thanks Dr. Ashis Kumar Nandy for the useful discussions.

\bibliographystyle{plain}
\bibliography{References}

\end{document}